\documentclass{jpp}

\usepackage{graphicx}
\usepackage{natbib}

\usepackage{epstopdf}
\usepackage{hyperref}
\hypersetup{
    bookmarks=true,         
    unicode=false,          
    pdftoolbar=true,        
    pdfmenubar=true,        
    pdffitwindow=false,     
    pdfstartview={FitH},    
    pdfsubject={Plasma fluid theory},   
    pdfproducer={GNU Make}, 
    pdfkeywords={keyword1} {key2} {key3}, 
    pdfnewwindow=true,      
    colorlinks=true,        
    linkcolor=blue,      
    citecolor=blue,         
    filecolor=blue,      
    urlcolor=blue           
}

\usepackage{graphics}
\usepackage{epsfig}
\usepackage{bm}
\usepackage{amsmath,amssymb}
\usepackage{stmaryrd}

\usepackage{relsize}

\usepackage{arydshln}
\usepackage{mathtools}
\usepackage{amsfonts}
\usepackage{amsmath}
\usepackage{amssymb}
\usepackage{graphicx}
\usepackage{bm}
\usepackage{dcolumn}
\usepackage{color}
\usepackage{comment} 
\usepackage{amstext,graphics,subfigure}

\usepackage{txfonts}

\def\bfB{\mathbf{B}}

\def\bfA{\mathbf{A}}
\def\bfa{\mathbf{a}}
\def\bfJ{\mathbf{J}}

\def\bfM{\mathbf{M}}

\def\bfL{\mathbf{L}}

\def\bfq{\mathbf{q}}
\def\bfv{\mathbf{v}}

\def\bfr{\mathbf{r}}

\def\bfeta{\boldsymbol{\eta}}
\def\bfPi{\boldsymbol{\Pi}}
\def\bfmu{\boldsymbol{\mu}}
\def\bftau{\boldsymbol{\tau}}
\def\bfxi{\boldsymbol{\xi}}

\def\bq{\begin{equation}}
\def\eq{\end{equation}}
\def\bqy{\begin{eqnarray}}
\def\eqy{\end{eqnarray}}

\def\bal#1\eal{\begin{align}#1\end{align}}

\def\de{\delta}

\def\ep{\epsilon}

\def\p{\partial}

\def\rh{\rho}
\def\si{\sigma}

\def\p{\partial}

\def\R{\mathbb{R}}

\def\calf{\mathcal{F}}

\def\calj{\mathcal{J}}

\def\call{\mathcal{L}}

\def\calt{\mathcal{T}}

 \def\q#1#2{q^{\hspace{1pt}#1}_{\,,\hspace{1pt}#2}}
 \def\a#1#2{a^{\hspace{1pt}#1}_{\,,\hspace{1pt}#2}}

\newcommand{\nc}{\newcommand}
\nc{\eqeqref}[1]{Eq.~\eqref{eq:#1}}
\nc{\eqseqref}[2]{Eqs.~\eqref{eq:#1}-\eqref{eq:#2} }
\nc{\secref}[1]{Sec.~\ref{sec:#1}}
\nc{\secsref}[2]{Sec.~\ref{sec:#1}-Sec.~\ref{sec:#2}}
\nc{\ssecref}[1]{Sec.~\ref{ssec:#1}}
\nc{\ssecsref}[2]{Sec.~\ref{ssec:#1}-Sec.~\ref{ssec:#2}}

  \shorttitle{Gyroviscous magnetohydrodynamic models}
  \shortauthor{M. Lingam, P. J. Morrison,  and A. Wurm}

\title{A class of three-dimensional gyroviscous magnetohydrodynamic models}

\author{Manasvi Lingam\aff{1,2} 
\corresp{\email{mlingam@fit.edu}},
Philip J. Morrison,\aff{3} \and Alexander Wurm\aff{4}}
\affiliation{\aff{1}Department of Aerospace, Physics and Space Sciences, Florida Institute of Technology, Melbourne FL 32901, USA \aff{2} Institute for Theory and Computation, Harvard University, Cambridge MA 02138, USA \aff{3}Department of Physics and Institute for Fusion Studies, The University of Texas at Austin, Austin, TX 78712, USA
\aff{4} Department of Physical and Biological Sciences, Western New England University, Springfield, MA 01119, USA}
 
\begin{document}

\maketitle

\begin{abstract}
A Hamiltonian and action principle formalism for deriving three-dimensional gyroviscous magnetohydrodynamic models is presented. The uniqueness of the approach in constructing the gyroviscous tensor from first principles and its ability to explain the origin of the gyromap and the gyroviscous terms are highlighted. The procedure allows for the specification of free functions, which can be used to generate a wide range of gyroviscous models. Through the process of reduction, the noncanonical Hamiltonian bracket is obtained and briefly analysed.

\end{abstract}

\section{Introduction}
The importance of finite Larmor radius (FLR) effects in plasma physics is well documented \citep{Brag58,RT62,RS65,braginskii65,Lil72,CQS87,HM92,Mik92,HW98,SP08,HD16,GKP19}. A broad class of models that incorporate FLR effects are those that fall under the fluid category, i.e., the momenta of the underlying particles are integrated out to yield mean field theories that describe the evolution of physical quantities such as density, fluid velocity, etc. The advantage of the fluid formalism stems from the fact that the complex dynamics of a multiparticle system is reduced to a few dynamical equations that are capable of accurately capturing its essential properties.

Fluid models that include FLR effects are often constructed by incorporating kinetic effects, e.g., by moving from particle phase-space coordinates to guiding centre coordinates \citep{HW83,HHM86,B92,SPH95,Be01}; models with FLR contributions incorporate kinetic effects of importance such as Landau damping and gyroradius averaging \citep{HDP92,BH96,SHD97,WSD97,SH01,SKW05,Mad13}. A second approach involves expansions in the smallness of the Larmor radius as compared to a characteristic length scale of the system and the imposition of closures for higher-order moments \citep{McM65,KG66,Bow71,PSH98,GPS05,SC06,ramos05a,ramos07,pass07,ramos10,ramos11,pass12,PST17,PHL17}. A third method uses the Hamiltonian framework to construct full and reduced magnetohydrodynamic (MHD) models endowed with FLR and other effects \citep{MH84,MCT84,HHM86,HHM87,B08,TMWG08,ICTC11,WT12,CGW13,LM14,l15,LiM15,PST18}. One of the chief advantages of Hamiltonian methods, as explained in the forthcoming sections, is that they are amenable to the extraction of naturally conserved quantities (the Casimirs) and analysing equilibria and stability.

The Hamiltonian formalism is deeply entwined with its twin approach, building models from an action principle - together, we will refer to them as the Hamiltonian and action principle (HAP) approach.\footnote{We intend this abbreviation to encompass all of the forms of action principles (Hamilton's principle, the phase space action, various constrained variational principles, etc.) and both canonical and noncanonical Hamiltonian descriptions. The HAP approaches of the present paper are Hamilton's principle yielding Lagrange's equations, which is here trivially related to the phase space action, the canonical Hamiltonian formulation in the Lagrange variable description, and the noncanonical Eulerian variable description.} The HAP formalism has a long history in fluid dynamics and plasma physics - examples of seminal publications prior to the 20th century include \citet{Lagrange,clebsch57,helmholtz58,clebsch59,hankel,kirchhoff76}.\footnote{Augustin-Louis Cauchy presented a Lagrangian formulation of three-dimensional incompressible hydrodynamics in a seminal, albeit forgotten, work in 1815 \citep{FV14}; see also \citet{FGV17}.} A summary of modern developments in this area can be found in the reviews by \citet{serrin59,TT60,SW68,Arn78,morrison82,HMRW,morrison98,AK98,morrison05,Ho08,morrison09,Ling15,SM16,morrison17,Tas17,Webb18}.

Using the action formalism has many advantages. For a starter, each term in the action has a clear physical meaning, which is not always the case when equations of motion have been derived using phenomenological or \emph{ad hoc} assumptions. Another advantage is that theories derived from action principles are naturally energy conserving. In some cases, equations of motion that had not been derived using the HAP formalism were erroneously believed to conserve energy (see e.g. \citealt{sc05,Sco07,tronci,kimura}). In addition, by performing an appropriate Legendre transformation, one can recover the Hamiltonian formalism, which is endowed with several advantages of its own. For a review of action principles in MHD models, we refer the reader to \citet{newcomb62,holm98,morrison09,Ling15,Webb18} and for the Hamiltonian formalism to \citet{morrison80,morrison82,HMRW,morrison98,morrison05,Tas17}. In particular, we mention its significance in studying symmetric MHD and its properties \citep{amp0,amp1,amp2a,amp2b}, and in constructing and analysing reduced MHD models \citep{MH84,HHM87,KPS94,KK04,WHM09,tassi10,TGP10,WT12,KWM15,TGB18,Tas19}.

Earlier we outlined different methods by which FLR effects can be incorporated into fluid models. It is worth noting that the Hamiltonian methods invoke the use of an interesting device - the gyromap, which was discovered in \citet{MCT84} and subsequently employed in the likes of \citet{HHM87} and \citet{ICTC11}. The gyromap is essentially a noncanonical transformation that maps the phase space to itself, and its chief advantage stems from the fact that it renders the noncanonical bracket of the gyroviscous MHD model identical to that of classical ideal MHD bracket \citep{morrison80} when expressed in terms of the new set of noncanonical variables; we will elaborate upon this point later in the paper.\footnote{The gyromap is a coordinate change from one set of dynamical variables to another. Its origin and usefulness will be expounded in Section \ref{Hamform}.} The origin of the gyromap was not properly understood until an action principle analysis in \citet{ling} was applied to a specific two-dimensional model, which assumed a particular ansatz for the internal energy and the gyromap. In this paper, we generalize the work of \citet{ling} to three dimensions, and present generic results in terms of freely specifiable functions. Furthermore, when we choose a particular ansatz for our FLR fluid model, we will use the physical principles of Larmor gyration to motivate the choice in detail. We will refer to this magnetofluid model as gyroviscous magnetohydrodynamics (GVMHD).

The paper is organized as follows. In Section \ref{Lagapp}, we outline the necessary tools for carrying out an action formulation of three-dimensional GVMHD. Then we proceed to build the action in Section \ref{APapproach}, where we motivate the reasoning behind the gyroviscous term. In Section \ref{EOMansatz}, the relevant equations of motion are presented and a particular choice of the gyroviscous ansatz is constructed. In Section \ref{Hamform}, we present the equivalent Hamiltonian formalism of this model. In Section \ref{gyrobracket}, we derive the gyroviscous MHD bracket and highlight the differences, compared with three-dimensional ideal MHD. Finally, we summarize our results in Section \ref{Conc}. Some of the salient auxiliary calculations are presented in the Appendices.

\section{The Lagrangian-variable approach to the action principle}
 \label{Lagapp}
 
In the first part of this section, we briefly describe Hamilton's principle of stationary action. In the second part, we highlight and outline the Lagrangian picture, and present a systematic methodology for moving to the more commonly used Eulerian picture.

\subsection{Hamilton's principle of stationary action}
\label{ActPrin}

The process involved in constructing the action for fluid models has been well-known since \citet{Lagrange}. Once the generalized coordinates  $q_k(t)$ are chosen, where $k$ runs over all possible degrees of freedom, the action is determined via
\bq
S[q] = \int^{t_1}_{t_0} \! dt\,L\left(q,\dot{q},t\right),
\eq
with $L$ representing the Lagrangian. It must be noted that $S$ is a ``functional", i.e., its domain and range are functions and real numbers respectively. Hamilton's principle states that that the equations of motion are the extrema of the action, i.e., we require $\delta S[q]/\delta q^k = 0$, where the functional derivative is defined as follows
 \bq
\de S[q;\de q]=
\left.\frac{d S[q +\ep \de q]}{d \ep}\right|_{\ep =0}
=: \left\langle \frac{\de S[q]}{\de q^{i}},\de q^{i}\right\rangle\, .
\eq
The continuum version is very similar to the discrete case since the discrete index $k$ is replaced by a continuous one, which we denote by $a$. The coordinate $q$ is a function of $a$ and $t$, and tracks the location of a fluid particle labelled by $a$. We also note the following important quantities which are used  throughout the paper: the deformation matrix $\p q^i/\p a^j=:\q ij$ and the corresponding determinant, the Jacobian, $\mathcal{J}:= \det(\q ij)$. The volume evolves in time via
\bq
d^3q=\mathcal{J} d^3a\,,
\label{vol}
\eq
and the area is governed by
\bq
(d^2q)_i=  \mathcal{J} \a ji \, (d^2a)_j\,,
\label{area}
\eq
where $\mathcal{J} \a ji$ is the transpose of the cofactor matrix of $\q ji$. The quantities and the relations introduced above can be used to generate a wide range of identities. One can find a detailed discussion of these, for example, in \citet{serrin59,morrison98,Ben06}.

\subsection{Two representations: the Lagrangian and the Eulerian points of view}
\label{ssec:AOLE}

The Lagrangian position $q$ evolves in time and is entirely characterized by its label $a$. But the fluid parcels are not solely determined by the position alone; they can also carry with them a certain density, entropy, and magnetic field. As the fluid moves along its trajectory, these quantities are also transported along with it, and are consequently characterized only by the label $a$ as well. We will refer to these quantities as {\it attributes}. As the label $a$ is independent of time, these attributes serve as Lagrangian constants of motion. The subscript $0$ will be used to label the attributes, in order to distinguish them from their Eulerian counterparts.

Let us now consider the Eulerian picture. All Eulerian fields depend on the position $\bfr:=(x^1,x^2,x^3)$ and time $t$, which can both be measured in the laboratory. As a result, we shall refer to these fields as {\it observables}. Moving from the Eulerian to Lagrangian viewpoint and \emph{vice versa} is accomplished with the Lagrange-Euler maps which we describe below in more detail.

The Eulerian velocity field $\bfv(\bfr,t)$ is the velocity of the fluid element at a location $\bfr$ and time $t$. If we seek to preserve the equivalence of the Lagrangian and Eulerian pictures, this must also equal $\dot{\bfq}(\bfa,t)$. As a result, it is evident that we require $\dot{\bfq}(\bfa,t)=\bfv(\bfr,t)$, where the dot indicates that the time derivative is obtained at fixed label $\bfa$. However, there is a discrepancy since the left-hand side is a function of $\bfa$ and $t$, while the right-hand side involves $\bfr$ and $t$. This conundrum is resolved by noting that the fluid element is at $\bfr$ in the Eulerian picture, and at $\bfq$ in the Lagrangian one. Hence, we note that $\bfr=\bfq(\bfa,t)$, which implies that $\bfa=\bfq^{-1}(\bfr,t)=:\bfa(\bfr,t)$ upon inversion. As a result, our final Lagrange-Euler map for the velocity is
\bq
\bfv(\bfr,t) =\left.\dot{\bfq}(\bfa,t)\right|_{\bfa=\bfa(\bfr,t)}\,.
\eq

Now we consider the attributes defined earlier, which we have noted are carried along by the fluid. The first attribute is the entropy of the fluid particle, which we shall label $s_0$. For ideal fluids, one expects the entropy to remain constant along the fluid trajectory. In other words, the Eulerian specific entropy $s(\bfr,t)$ must also remain constant throughout, implying that $s=s_0$.  Apart from entropy, the magnetic stream function $\psi$ for  two-dimensional gyroviscous MHD \citep{amp2a,ling} also obeys this property.

Next, we can consider attributes which obey a conservation law similar to the density. The conservation law in this case is that of mass conservation. The attribute is denoted by $\rho_0(\bfa)$ and the observable by $\rh(\bfr,t)$. The statement of mass conservation in a given (infinitesimal) volume amounts to 
\ $\rho(\bfr,t)d^3r=\rho_0(\bfa)d^3a$.  Using Eq.~(\ref{vol}) we obtain $\rho_0=\rho \calj$. As a result, we have found the Lagrange-Euler map for $\rho$. There exist other attribute-observable pairs in the literature, which also possess similar conservation laws, such as the entropy density.

In the case of magnetofluid models, it is often advantageous to introduce the magnetic field attribute $\bfB_0(\bfa)$. In the case of ideal magnetofluid models, the conservation law of frozen-in magnetic flux is applicable. In algebraic terms, this amounts to $\bfB\cdot \mathbf{d^2r}=\bfB_0\cdot \mathbf{d^2a}$, and from Eq.~(\ref{area}) we obtain $\calj B^i=\q ij  \,B_0^j$.

In all of the above expressions, the picture is still incomplete since we need to remove the $\bfa$-dependence of the attributes. In a manner similar to that undertaken for the velocity, we evaluate the attributes at $\bfa=\bfq^{-1}(\bfr,t)=:\bfa(\bfr,t)$. This completes our prescription, and one can fully determine the observables once we are provided the attributes in conjunction with the Lagrangian coordinate $\bfq$.

We may also represent the Lagrange-Euler map in an integral form, which permits a more intuitive interpretation. We shall start with the assumption that the attribute-observable relations are found via appropriate conservation laws. We have stated before that one moves from the Lagrangian to the Eulerian picture by `plucking out' the fluid element that happens to be at the Eulerian observation point $r$ at time $t$. Such a process is accomplished mathematically via the delta function $\delta(\bfr-\bfq(\bfa,t))$. For instance, we see that the density can be treated as follows:
\bqy
\rh({\bfr},t)&=&\int_D \!d^3a
\, \rh_0(\bfa) \, \de\left({\bfr}-{\bfq}\left(\bfa,t\right)\right)
\nonumber\\
&=&\left. \frac{\rh_0}{\mathcal{J}}\right|_{\bfa=\bfa({\bfr},t)}\,.
\label{rhoEL}
\eqy
Further below, we will also use a new variable, the canonical  momentum density  $\bfM^c=(M^c_1,M^c_2,M^c_3)$, which is related to its Lagrangian counterpart via
\bqy
\bfM^c(\bfr,t)&=&\int_D \!d^3a \,
{\bfPi}(\bfa,t) \, \de\left({\bfr}-{\bfq}(\bfa,t)\right)
\nonumber\\
&=& \left.
\frac{\bfPi(\bfa,t)}{\mathcal{J}}
\right|_{\bfa=\bfa(\bfr,t)}\,.
\label{canMomEL}
\eqy
 For ideal MHD,  the canonical momentum density is $\bfPi(a,t)=(\Pi_1,\Pi_2,\Pi_3)=\rh_0 \dot{\bfq}$.  It is worth noting that $\bfPi(\bfa,t)$ can be found from the Lagrangian through $\bfPi(\bfa,t) = {\de L}/{\de \dot{\bfq} }$ and does not necessarily equal  $\rh_0 \dot{\bfq}$ in general. One can also construct such integral relations for the entropy and the magnetic field. We refer the reader to \citet{ling} for a more detailed discussion along these lines.

\section{Action principle for a generic magnetofluid}
\label{APapproach}

The first part of this section is devoted to a brief description of the procedure outlined in \citet{morrison09,ling} for constructing action principles for magnetofluid models. Some of the advantages have been highlighted in the introduction, and others can be found in, for example, \citet{morrison09,ling}. Then, we proceed to construct our action and motivate our choice of terms along the way.

\subsection{The general action}
\label{process}

The domain of integration $D$ is chosen to be a subset of $\R^{3}$. Central to our formulation is the Lagrangian coordinate $\bfq\colon D\rightarrow D$, which we shall assume to be a well-behaved function with the required smoothness, invertibility, etc. Next we need to specify our set of observables, or alternatively our set of attributes. For our models, we work with $\mathfrak{E}=\{\bfv,\rho,\sigma,\bfB\}$ where $\sigma = \rho s$ is the entropy density. Finally, we shall impose the {\it Eulerian closure principle}, which is necessary for our model to be  `Eulerianizable.'\ \ Mathematically, this principle amounts to the action being fully expressible in terms of the Eulerian observables. Physically, the principle states that our theory must be solely describable in terms of physically meaningful quantities, the observables, and must also give rise to equations of motion in terms of these observables. As a result, we require our action to be given via
\bq
S[\bfq]:= \int_{T} dt\int_D d^3a\,\call\left(\bfq,\dot{\bfq},\p \bfq/\p \bfa\right) =: \bar{S}\left[\mathfrak{E}\right].
\label{Lagaction}
\eq
As per the Eulerian closure principle, this amounts to finding an action $\bar{S} = \int_\calt dt \int_D d^3r\,\bar{\call}$ in terms of the Eulerian observables. The presence of the bar indicates that the action and the Lagrangian density are expressed solely in terms of the observables.

\subsection{Constructing the gyroviscous action} 
\label{Gyroact}

The first step in the process involves the construction of the kinetic energy, which must also satisfy the closure principle. Using the analogy with particle mechanics, we know that it equals
\bq
S_{\mathrm{kin}}:= \int_\calt dt\int_D d^3a\, \frac{1}{2} \rh_0 |\dot{\bfq}|^2 = \int_\calt dt\int_D d^3r\, \frac{1}{2} \rh |\bfv|^2,
\eq
where the last equality is obtained by using relations outlined in Section \ref{ssec:AOLE}.

The internal energy per unit mass is a function of the entropy density and the density, and in Eulerian terms it can be represented by $U\left(\rh,\sigma\right)$. Using the inverse Lagrange-Euler maps, we can construct the Lagrangian internal energy density accordingly.
\begin{eqnarray}
S_{\mathrm{int}}&:=& \int_\calt dt\int_D d^3a\, \rh_0 U\left(\frac{\rh_0}{\calj},\frac{\sigma_0}{\calj}\right)\nonumber\\
 &\,=& \int_\calt dt\int_D d^3r\, \rho U\left(\rh,\sigma\right),\label{Sint}
\end{eqnarray}

The next step is the construction of the magnetic energy, and we use the same process outlined for the internal energy, viz. we determine the Eulerian term and obtain the Lagrangian version consequently through the Lagrange-Euler map.
\begin{eqnarray}
S_{\mathrm{mag}}&:=& \int_\calt \int_D d^3r\, \frac{1}{2} |\bfB|^2,\nonumber\\
&\,=& \int_\calt dt\int_D d^3a\, \frac{1}{2 \calj} \q ij \, \q ik \, B_0^j B_0^k\, . \label{Smag}
\end{eqnarray}
The magnetic energy is actually $|\bfB|^2/8\pi$ in CGS units but we drop the factor of $4\pi$ henceforth by scaling it away through the adoption of Alfv\'enic units.

Now we are ready to construct the most important term which will be responsible for the gyroviscosity. The gyroviscous term is taken to be \emph{linear} in $\dot{\bfq}$ and is given by
\begin{equation}
\label{Sgyro}
S_{\mathrm{gyro}} = \int_\calt dt\int_D d^3a\, \dot{\bfq} \cdot \bfPi^{\star} = \int_\calt dt\int_D d^3r\, \bfv \cdot \bfM^\star.
\end{equation}
In other words, we operate under the premise that $\bfPi^{\star}$ is solely a functional of $\bfq$ and $t$. As the Eulerian perspective is inherently endowed with physical variables (e.g., density and magnetic field), we will focus on the Eulerian equivalent of $\bfPi^{\star}$; from the Eulerian closure principle we obtain the relation
\bq
\bfM^\star = \left.
\frac{\bfPi^\star(\bfa,t)}{\mathcal{J}}
\right|_{\bfa=\bfa(\bfr,t)}\,.
\label{Mstarthree-dimensional}
\eq
The complete action functional is now given by
\begin{equation} \label{finalaction}
S = S_{\mathrm{kin}} - S_{\mathrm{int}} - S_{\mathrm{mag}} + S_{\mathrm{gyro}}.
\end{equation}
The action of \eqref{finalaction} is  general, but not the most general second-order (in $\bfv$) action that satisfies the Eulerian closure principle.  For example, the term $S_{\mathrm{kin}}$ could be generalized by replacing its integrand with  $\rho_0G |\dot{\bfq}|^2/2|_\bfa = \rho G(\rho, \si, \bfB)|\bfv|^2/2$ and  the integrand of $S_{\mathrm{int}}$  could be replaced by  $\rho_0 U|_\bfa= \rho U(\rho, \si, \bfB)$, a form that was shown in \citet{morrison82} to allow for anisotropic pressure. Here both $G$ and $U$ could be arbitrary functionals (including derivatives) of their arguments.  Similarly the term $S_{\mathrm{mag}}$ could be generalized.

The Eulerian canonical momentum density is defined via \eqref{canMomEL}, which can be computed by finding the Lagrangian canonical momentum using  $\bfPi(\bfa,t) = {\de L}/{\de \dot{\bfq} }$ and Eulerianizing it. Upon doing so, we arrive at the so-called gyromap, a device introduced in \citet{MCT84} as follows:
\begin{equation}\label{gyroM}
    \bfM^c := \rh \bfv + \bfM^\star =\bfM + \bfM^\star\,.
\end{equation}
The benefit of employing the gyromap and its natural origin will be discussed in Section \ref{Hamform} and further explicated in Appendix \ref{AppB}.

So far we have only required $\bfM^\star$ to satisfy the closure principle, i.e., that  it be expressible in terms of the subset $\{\rho,\sigma,\bfB\}\subset\mathfrak{E}$, including all possible Eulerian derivatives.  Given that $\bfM^\star$  is a momentum density, arising perhaps from underlying gyration of particles, a natural assumption is that it has the magnetization form
\bq
\bfM^\star = \nabla \times \bfL^\star\,, 
\label{helm}
\eq
i.e., we assume that $\bfM^\star$ is divergence-free. Since we are interested in a gyroviscosity due to gyromotion, this is a physically reasonable assumption.  However, one could replace \eqref{helm} by a Helmholtz decomposition for a more general collisionless viscosity.  The present choice is also motivated in part by the realization in \citet{MCT84} and \citet{ling} that this choice is consistent with existing two-dimensional gyroviscous models.  Because  $\bfM^\star$ has the units of momentum density, from which we see that the quantity $\bfJ^\star \propto (q/m) \bfM^\star$ resembles a current density. If one assumes that the fluid ``particles'' possess a finite magnetic moment, it follows that  the fluid must have a finite magnetization. In other words, one may identify $\bfJ^\star$ with the magnetization current density, which is divergence-free \citep{Jack98} and the current through an area depends on flux through a bounding curve. 
Are other choices possible and do any of them  conserve angular momentum?
Perhaps an even simpler way of envisioning the ansatz for $\bfM^\star$ is that it must emerge from the gyration of particles. In pictorial terms, this gyration is reminiscent of the effect generated by the curl of a vector field, which motivates our choice of $\bfM^\star$. Further grounds for assuming this particular expression are described in \citet[Section 5]{ling}. 
With this ansatz, evidently $\nabla \cdot \bfM^c = \nabla \cdot \bfM$, since the second term vanishes. Note that the RHS of this expression appears in the continuity equation, and we see that one could also replace it by the LHS if we operate with $\bfM^\star = \nabla \times \bfL^\star$. Furthermore, dimensional analysis permits the identification of $\bfL^\star$ with the angular momentum density. 

As we have reduced the question of determining $\bfL^\star$, we must ask ourselves as to whether any further simplifications are feasible. Once again, we can resort to physical intuition to gain an idea of what $\bfL^\star$ might look like. Without further special assumptions about the fluid, e.g., it having some intrinsic or extrinsic direction,  the vectorial character of $\bfL^\star$ must come from $\bfB$ or from the set of gradients of the observables; these and their cross products are the only vectors available. Thus, for example, a general form for $\bfL^\star$ could be composed of a linear combination of these vectors with coefficients dependent on $\rho, \si$, and $|\bfB|$. If we assume $\bfL^\star$ constitutes an internal angular momentum density of some kind associated with particle gyration, then it is reasonable to posit that it would tend to align with the magnetic field $\bfB$. Moreover, in the limit of a large magnetic field, the corresponding gyroradii would become small, owing to which the fluid particle may not be significantly affected by gradients on these scales. Combining the preceding arguments leads to the generic form
\bq
\bfL^\star= \calf\left(\rh,\sigma,|\bfB|\right) \bfB\,.
\label{genLform}
\eq
In Section \ref{originGyro} we will argue for further specification of the properties of \eqref{genLform}.

With the choice of \eqref{genLform}, the gyroviscous term of the action, expressed in terms of the observables is given by
\bqy
S_{\mathrm{gyro}} &=& \int_\calt dt\int_D d^3r\, \bfv \cdot \nabla \times \left[\bfB \calf\left(\rh,\sigma,|\bfB|\right) \right] \nonumber \\
&=& \int_\calt dt\int_D d^3r\, \calf \bfB\cdot \nabla \times \mathbf{v},
\eqy
where the second equality follows from integrating by parts and neglecting the boundary term. We shall use the latter operation consistently throughout the rest of the paper. Now that we have constructed the gyroviscous term, we note that it is still generic since there is considerable freedom in the choice of $\calf$.

\section{The equations of motion and the choice of ansatz}
\label{EOMansatz}

In this section, we shall present the equations of motion and discuss the origin of the gyroviscous terms, 
and why a specific choice of the free function $\calf$ emerges in a natural manner.

\subsection{The equations of motion} 
\label{EOM}

The equations for the density, entropy density, and the magnetic field can be determined via the attributes-observables relations defined through the appropriate conservation laws and the Lagrange-Euler maps. The entropy density and the density obey similar laws, given by
\begin{equation} \label{rhoEOM}
\frac{\p \rh}{\p t} + \nabla \cdot \left( \rh \bfv\right) = 0,
\end{equation}
\begin{equation} \label{sigmaEOM}
\frac{\p \sigma}{\p t} + \nabla \cdot \left( \sigma \bfv\right) = 0.
\end{equation}
The equation governing the magnetic field is
\begin{equation} \label{BEOM}
\frac{\p \bfB}{\p t} + \bfB \left(\nabla \cdot \bfv\right) - \left(\bfB \cdot \nabla\right) \bfv + \left(\bfv \cdot \nabla\right) \bfB = 0,
\end{equation}
which can be recast into the more familiar induction equation if $\nabla \cdot \bfB = 0$ is satisfied. If the constraint is obeyed, then we obtain
\begin{equation}
\frac{\p \bfB}{\p t} = \nabla \times \left(\bfv \times \bfB\right).
\end{equation}

The dynamical equation for the momentum is derived from $\delta S = 0$, and is thus equal to
\bqy \label{theEOM}
&&\frac{\partial}{\partial t}\left(\rho v^{k}\right) + \partial_j \left[\rho v^j v^k + \left(p+\frac{|\bfB|^2}{2}\right) \delta^{jk} - B^j B^k\right] \nonumber \\
&&-\partial_j \left[\bfB\cdot\left(\nabla\times \bfv\right)\left(\rho\frac{\partial{\cal F}}{\partial\rho}+\sigma\frac{\partial{\cal F}}{\partial\sigma}+|\bfB|\frac{\partial{\cal F}}{\partial|\bfB|}-\calf \right)\delta^{jk}\right] + \partial_j \left[\bfB\cdot\left(\nabla\times \bfv\right)\left(\frac{\partial{\cal F}}{\partial|\bfB|}\frac{B^{j}B^{k}}{|\bfB|}\right)\right] \nonumber \\
&&+ \partial_j \left[\epsilon_{kji}B^{i}\left(\frac{\partial{\cal F}}{\partial\rho}\frac{\partial\rho}{\partial t}+\frac{\partial{\cal F}}{\partial\sigma}\frac{\partial\sigma}{\partial t}+\frac{\partial{\cal F}}{\partial|\bfB|}\frac{\bfB}{|\bfB|}\cdot\frac{\partial \bfB}{\partial t}\right)\right] + \partial_j \left[\epsilon_{kji}{\cal F}\frac{\partial B^{i}}{\partial t}\right] \nonumber \\
&&-\partial_{j}\left[\epsilon_{lji}{\cal F}B^{l}\left(\partial_{k}v^{i}\right)-\epsilon_{kli}{\cal F}B^{j}\left(\partial_{l}v^{i}\right)\right] = 0,
\eqy
where repeated indices indicate summation (as per the Einstein convention), and we have employed the standard relationship between the internal energy and the scalar pressure $p$. We note that Eq.~(\ref{theEOM}) can be obtained in two different ways from the action. The first is to follow the conventional variation with respect to $q$ and obtain it accordingly. The second method involves the use of the procedure outlined in \citet{FR60,newcomb62} and is described in Appendix \ref{AppA}. For our model, Eqs.~(\ref{rhoEOM}), Eq.~(\ref{sigmaEOM}), Eq.~(\ref{BEOM}), and Eq.~(\ref{theEOM}) constitute the complete set of dynamical equations.

Before discussing the ansatz in more detail, a few observations regarding Eq.~(\ref{theEOM}) are in order. The second term occurring in the first line of this equation represents the ideal MHD momentum flux (enclosed in square brackets), which is seen from the absence of $\calf$ in it. The second line contain terms that are purely symmetric under the interchange $k \leftrightarrow j$. The third line contains terms that are wholly antisymmetric under $k \leftrightarrow j$. The fourth (and final) line contains terms that are neither purely symmetric nor purely antisymmetric. As a result, we see that the entire momentum flux tensor is not symmetric, as opposed to the ideal MHD tensor, or the two-dimensional gyroviscous tensor for the specific model considered in \citet{ling}. Note that we refer to the terms from line two onwards as {\it gyroviscous} because they are expressed in terms of the velocity shear, akin to viscous hydrodynamics. The gyroviscous tensor thus obtained above can be compared against the general expression(s) presented in \citet{ramos05b}. Furthermore, these effects arise from charged particle gyration - the latter aspect is explored below.

\subsection{The origin of the gyroviscous ansatz}
 \label{originGyro}

In Section \ref{Gyroact}, we briefly the process involved in constructing a generic gyroviscous term. Now, we shall draw upon further physics to select a \emph{specific} choice for the ansatz.

First, let us suppose that we start out with the notion of an internal angular momentum $\bfL^\star$. In order to understand where this angular momentum originates, we recall an identity from electromagnetism which relates the angular momentum to the magnetic moment via the gyromagnetic ratio, $(2m)/e$. If we consider a two species model of ions and electrons, then the ions will play the dominant role, owing to their higher mass. Hence, we know that $\bfL^\star = \frac{2m}{e} \bfmu$. The magnetic moment $\bfmu$ is typically an adiabatic invariant in plasmas, and its magnitude is given by $|\bfmu|=\frac{m \bfv_\perp^2}{2|\bfB|}$, which is proportional to $P_\perp/|\bfB|$ where $P_\perp$ denotes the perpendicular component of the (anisotropic) pressure. But, the magnetic moment is a vector and the most natural way to construct a vector is through the unit vector of the magnetic field. Putting these results together, we find that a natural ansatz (albeit a specific one) for $\bfL^\star$ is given by
\begin{equation}\label{Lsdef}
\bfL^\star = \alpha \frac{m}{2e} \frac{P_\perp}{|\bfB|^2} \bfB,
\end{equation}
where $\alpha$ is a dimensionless proportionality constant, which can be arbitrarily specified; in the ensuing analysis, we set $\alpha = 1$ for simplicity. By comparison with the more general ansatz outlined in Section \ref{Gyroact}, we find that they are identical when $\calf = \alpha \frac{m}{2e} P_\perp/|\bfB|^2$.

The function $P_\perp$ is a function of $\sigma$, $\rho$ and $|\bfB|$. For a more detailed discussion of the anisotropic pressure, we refer the reader to \citet{kimura}. It is defined as
\begin{equation} \label{Pperp}
P_\perp = \rho^2 \frac{\p U}{\p \rh} + \rh |\bfB| \frac{\p U}{\p |\bfB|},
\end{equation}
an expression that first appeared in \citet{morrison82}, where $U$ is the internal energy that is a function of $\rh$ and $\sigma$, but also of the magnetic field; see also \citet{HMM13}. If we wish to forgo anisotropy, then we assume that $U$ is independent of $B$, and hence the second term in the above term vanishes. This assumption was used in deriving the equation of motion Eq.~(\ref{theEOM}) since the internal energy introduced in Eq.~(\ref{Sint}) had no $\bfB$-dependence. Such an assumption also leads to the pressure tensor becoming isotropic, given by the first term of Eq.~(\ref{Pperp}) alone.

In summary, the ansatz constructed was chosen such that the gyroviscosity (and consequently the momentum transport) arises via the gyration of charged particles, thereby lending the term its name. The fact that momentum transport could take place via such gyrations was first noted by \citet{CC53,kaufman60} in the 1950s and 1960s. This principle was applied to incompressible gyrofluids in \citet{newcomb72,newcomb73,newcomb83} and compressible gyrofluids in \citet{morrison09,ling}, who showed that this specific ansatz yielded results that were fully compatible with the two-dimensional version of the Braginskii tensor \citep{braginskii65}. 

Lastly, we note that substituting (\ref{Lsdef}) in (\ref{gyroM}) after employing $\bfM^\star = \nabla \times \bfL^\star$ will yield a number of extra terms with the same dimensions as $\bfM = \rho \bfv$. Hence, if one divides the expression throughout by $\rho$, the contributions arising from $\bfM^\star$ have the dimensions of velocity and possess physical interpretations. The first term, which is proportional to $\left(\bfB \times \nabla P_\perp\right)/|\bfB|^2$, amounts to the diamagnetic drift velocity. The second term, which is proportional to $P/|\bfB|^3 \left(B \times \nabla |\bfB|\right)$ is analogous to the $\nabla |\bfB|$ drift velocity for charged particles. This correspondence has been pointed out in \citet[Section 6]{MCT84}.
 
\section{Angular momentum conservation and its ramifications}
 \label{AngMom}
 
In this section, we discuss the chief unusual property of our model - the lack of an `orthodox' angular momentum conservation, and its resolution. We also present a brief illustration of its ramifications in an astrophysical context.

\subsection{Constructing a hybrid conserved angular momentum}
 \label{HybridAngMom}

When we perform the constrained variation of our action, we recover
\begin{eqnarray} \label{momcons}
\frac{\partial M^c_i}{\partial t} &+& \partial_j T_{ij} = 0, \nonumber \\
T_{ij} &=& M^c_i v_j - \frac{\partial \mathcal{L}}{\partial \left(\partial_j v_k\right)} \left(\partial_i v_k\right) + \frac{\partial \mathcal{L}}{\partial B^i} B_j + \delta_{ij} \left[\mathcal{L} - \frac{\partial \mathcal{L}}{\partial B^k} B^k  - \frac{\partial \mathcal{L}}{\partial \rho}\, \rho \right].
\end{eqnarray}
Additional details can be found in \citet[Equations 7.6-7.8]{holm98} and \citet[Section 3]{LM14}. Note that the Lagrangian density $\mathcal{L}$ in the above expression refers to the one present in Eq.~(\ref{finalaction}). A rather unusual fact emerges if one inspects the above energy-momentum tensor: when one considers ideal MHD, or even Hall and extended MHD, the tensor $T_{ij}$ is symmetric. In turn, this ensures that the angular momentum $\bfM={\bfr} \times  \rho \bf{v}$ is conserved. However, this is evidently not the case for the above energy-momentum tensor. 

This fact is not unusual because a number of hydrodynamic models are known to possess asymmetric energy-momentum tensors. In particular, if the constituent `particles' (which may be fluid parcels) have an internal degree of freedom (i.e., spin), the energy-momentum tensor of the fluid will manifest a non-symmetric component \citep{Pap49,SL67,OS76,Dew77,Eva79,Kop90,Lin15}. Examples of hydrodynamic models with asymmetric energy-momentum tensors include ferrohydrodynamics \citep{Ros85,Bil05} and nematics \citep{DP93}. Although many core plasma models are characterized by symmetric energy-momentum tensors \citep{PM85,Smil85}, other plasma models feature asymmetric energy-momentum tensors \citep[e.g.,][]{AJB10}. In consequence, not all components of the angular momentum will be conserved, although the toroidal component is conserved in such models \citep{SS10}.

To resolve this, we will adopt the procedure delineated in \citet{McL66}. We begin with the observation that the first expression in Eq.~(\ref{momcons}) remains invariant under the transformations $M^c_i \rightarrow M^c_i + \partial_j {\Sigma}_{ij}$ and $T_{ij} \rightarrow T_{ij} - \partial {\Sigma}_{ij}/\partial t$. Let us suppose that we choose $\partial {\Sigma}_{ij}/\partial t$ to be the antisymmetric part of $T_{ij}$, thereby ensuring that $T_{ij} - \partial {\Sigma}_{ij}/\partial t$ is purely symmetric. Hence, by utilizing this choice of ${\Sigma}_{ij}$, we find that
\begin{equation} \label{antisym}
\frac{\partial {\Sigma}_{ij}}{\partial t} = T^A_{ij} = \frac{1}{2} \left(T_{ij} - T_{ji}\right) = \epsilon_{ijk} \tau_k,
\end{equation}
where $\bf{\tau}$ has the units of torque density and is given by
\begin{equation} \label{torque}
\tau_k = \frac{1}{2}\left[ \epsilon_{kab} M^c_a v_b + \frac{m}{2e} \frac{P_\perp}{|\bfB|^2} \left(B_k \partial_l v_l - B_l \partial_k v_l\right) \right].
\end{equation}
The first term in the above expression is ${{\bfM}}^c\times{\bfv}$, which can also be expressed as ${{\bfM}}^\star \times{\bfv}$ since $\rho {{\bfv} \times {\bfv}} = 0$. The second and third terms are proportional to $\left(\nabla \cdot {{\bfv}}\right) {\bfB}$ and $\left(\nabla {\bfv}\right) \cdot {\bfB}$ respectively. Since we know that $\tau$ behaves as a torque density, let us define a dynamical variable $\mathcal{S}$ such that $\partial \mathcal{S}_k/ \partial t = \tau_k$; this constitutes a relation that mirrors the conventional torque-angular momentum relation in classical mechanics. Using this in Eq.~(\ref{antisym}), we find that $\Sigma_{ij} =  \epsilon_{ijk} \mathcal{S}_k$. With these ingredients, we can now construct a symmetric momentum conservation law as follows:
\begin{equation}
\frac{\partial M^{tot}_i}{\partial t} + \partial_j T^S_{ij} = 0,
\end{equation}
with $T^S_{ij}$ representing the symmetric energy-momentum tensor and $M^{tot}_i = M^c_i + \epsilon_{ijk} \partial_j \mathcal{S}_k$. As the resultant energy-momentum tensor is symmetric, it follows that the corresponding angular momentum ${{\bfr}} \times {{\bfM}}^{tot}$ is conserved. 

The ramifications of $\mathcal{S}$ are manifold. It can be interpreted as an intrinsic angular momentum density generated from the torque density Eq.~(\ref{torque}). This is consistent with prior works \citep{Pap49,SL67,OS76,Dew77,Eva79,Kop90} that outlined the connections between intrinsic angular momentum and a non-symmetric energy-momentum tensor. A second justification arises from ${{\bfM}}^{tot} = {\bfM}^c + \nabla \times \mathcal{S}$, implying by dimensional analysis that $\mathcal{S}$ has the dimensions of angular momentum density. If we define ${\bfM}^{int} = \nabla \times \mathcal{S}$, we see that $\nabla \cdot {\bfM}^{int} = 0$. The kinship between ${\bfM}^\star$ and ${\bfM}^{int}$ is obvious as they are both generated via an internal angular momentum mechanism and are divergence-free.

Let us now summarize our results. We defined a dynamical variable $\mathcal{S}$ such that it obeys $\partial \mathcal{S}_i/\partial t = \tau_i$ where $\tau$ is given by Eq.~(\ref{torque}), and it emerges from the antisymmetric part of the original energy-momentum tensor. We also find that the new momentum ${\bfM}^{tot} = {\bfM} +  \left({\bfM}^\star + {\bfM}^{int}\right)$ yields a symmetric momentum tensor (which is the symmetric part of the old one). Using the expressions for ${\bfM}^\star$ and ${\bfM}^{int}$, we have 
\begin{equation}
{\bfM}^\star + {\bfM}^{int} = \nabla \times \left({\bfL}^\star + \mathcal{S}\right).
\end{equation}
Hence, we can define a composite intrinsic angular momentum ${\bfJ} = {\bfL}^\star + \mathcal{S}$, akin to the total angular momentum in quantum mechanics \citep{Wein15}. The introduction of ${\bfJ}$ yields ${\bfM}^{tot} = {\bfM} + \nabla \times {\bfJ} $, which is simple in form and has an immediate physical interpretation. The angular momentum corresponding to ${\bfM}^{tot}$ is conserved, and is given by ${\bfr} \times {\bfM}^{tot}$. Hence, the total angular momentum defined below is an invariant. 
\begin{equation} 
\label{AMHybrid}
\int d^3r\,\left[{\bfr} \times {\bfM} + {\bfr}\times \left(\nabla \times {\bfJ}\right)\right].
\end{equation}

Before proceeding further, some major aspects concerning the two-dimensional  GVMHD model described in \citet{MCT84} and \citet{ling} merit further explication. To begin with, we can rewrite Eq.~(\ref{momcons}) as follows:
\begin{equation}\label{Momconstwo-dimensional}
    \frac{\partial M_i}{\partial t} + \partial_j \tilde{T}_{ij} + \left[ \frac{\partial M^\star_i}{\partial t} + \partial_j \left(M^\star_i v_j \right) \right] = 0
\end{equation}
where we have introduced the new energy-momentum tensor
\begin{equation}\label{EMTenstwo-dimensional}
    \tilde{T}_{ij} = \rho v_i v_j - \frac{\partial \mathcal{L}}{\partial \left(\partial_j v_k\right)} \left(\partial_i v_k\right) + \frac{\partial \mathcal{L}}{\partial B_i} B_j + \delta_{ij} \left[\mathcal{L} - \frac{\partial \mathcal{L}}{\partial B_k} B_k  - \frac{\partial \mathcal{L}}{\partial \rho}. \rho \right].
\end{equation}
The first key point worth highlighting here is that \citet{MCT84,ling} adopted: (i) a specific equation-of-state (EOS) for $P_\perp$ wherein $P_\perp/|\bfB|$ was a Lie-dragged scalar density, and (ii) the choice $\bfB = B_z \hat{z}$ for the magnetic field. These two conditions collectively ensured that $\bfL^\star$ had only one component and that the components of $\bfM^\star$ behaved as scalar densities that underwent Lie-dragging; in other words, the term inside the square brackets of Eq.~(\ref{Momconstwo-dimensional}) vanishes identically for the two-dimensional GVMHD model.

The second essential point is that two-dimensional GVMHD did not include any variables that were Lie-dragged as vector densities of rank unity. In contrast, the magnetic field in three-dimensional MHD and GVMHD plays this role \citep{morrison82,LM14},\footnote{Alternatively, if one considers the Hodge dual of the magnetic field, it constitutes an example of a Lie-dragged two-form \citep{TY93}.} but $B_z$ in two-dimensional GVMHD is a Lie-dragged \emph{scalar} density as seen from \citet[Equation 3]{MCT84}; to put it differently, $B_z$ in two-dimensional GVMHD is advected the same way as the plasma density $\rho$. Thus, the terms in Eq.~(\ref{EMTenstwo-dimensional}) involving $B_i$'s are rendered irrelevant because they were derived under the assumption that the magnetic field is a Lie-dragged vector density. Hence, these two facts collectively ensure that the only potential source of asymmetry in the energy-momentum tensor of two-dimensional GVMHD is the second term on the RHS of Eq.~(\ref{EMTenstwo-dimensional}). When one utilizes the particular EOS for this model in conjunction with $M_z = 0$ and $\bfB = B_z \hat{z}$, it can be shown \citep{MCT84,ling} that the gyroviscous term of two-dimensional GVMHD yields the contribution
\begin{equation}
 {T}_{ij}^{(\mathrm{two-dimensional-GV})} = \frac{m}{2e} \frac{P_\perp}{B_z} \mathcal{N}_{jlik} \partial_k v_l\,, \quad \quad \mathcal{N}_{jlik} = \delta_{jk} \epsilon_{li} - \delta_{li} \epsilon_{jk}
\end{equation}
to the energy-momentum tensor, which turns out to be fully symmetric. 

The above discussion serves to illustrate how and why the energy-momentum tensor of the simplified two-dimensional GVMHD model of \citet{MCT84,ling} is symmetric in nature. However, in order to achieve this symmetry, a number of restrictions on the equation-of-state as well as the magnetic field and momentum density had to be imposed. When all of these constraints are relaxed, which is the case for three-dimensional GVMHD, one finds that an asymmetric energy-momentum tensor is obtained.

\subsection{An illustration of the formalism}

We have already noted earlier that the kinetic angular momentum ${ r} \times {\bfM}$ is not conserved. However, we have seen that the angular momentum described in Eq.~(\ref{AMHybrid}) is conserved. Together, these imply that the rate of loss (or gain) of the kinetic angular momentum ${\bfr} \times {\bfM}$ is precisely equal to the rate of gain (or loss) of the intrinsic angular momentum ${\bfJ}$. Let us recall that $\mathcal{S}$ comprises a part of ${\bfJ}$, and we know that $\partial \mathcal{S}_i/\partial t = \tau_i$ where $\bftau$ is given by Eq.~(\ref{torque}). The first term in Eq.~(\ref{torque}) reduces to ${\bfM}^\star \times {\bfv}$, as noted earlier. It is worth mentioning that the additional two terms are quite different, but exhibit a similar scaling. Hence, we shall use only the first term in our subsequent analysis. The total torque (denoted by $\tilde{\mathcal{T}}$) is found by integrating this term over the volume, and thus gives rise to the scaling
\begin{equation} 
\label{torquescal}
\tilde{\mathcal{T}} = \int {\bfM}^\star \times \bfv\,d^3r \sim \frac{m}{e} \left(\frac{P_\perp}{|\bfB|}\right) \Omega R^3,
\end{equation}
where we have dropped the numerical factors and used a characteristic velocity of $\Omega R$, with $R$ denoting the radius of the (spherical) object. It is evident that the scaling will be entirely determined by the EOS that is adopted.

Next, let us evaluate the spin-down rate, by using the relation $\tilde{\mathcal{T}} = I \dot{\Omega}$, from classical mechanics. The moment of inertia, dropping all numerical factors, is approximately $M R^2 \sim \rho R^5$. Using this in Eq.~(\ref{torquescal}), we find that
\begin{equation} \label{spindown}
\dot{\Omega} \sim \frac{m}{e} \left( \frac{P_\perp}{\rho |\bfB|} \right) \Omega R^{-2}.
\end{equation}
The above relation indicates that $\dot{\Omega} \propto \Omega$ (holding other quantities fixed). The EOS depends only on $\rho$, $s$ and $|\bfB|$ and hence we can conclude that the relation $\dot{\Omega} \propto \Omega$ is likely to be independent of the choice of the EOS. If we treat $\rho$ and $R$ to be independent variables, i.e.\  by choosing $M$ to be the dependent variable, one can also conclude that $\dot{\Omega} \propto R^{-2}$ will be independent of the EOS. The characteristic time $t_c = \Omega/\dot{\Omega}$, is expected to be independent of $\Omega$ and is given by
\begin{equation} \label{tchar}
t_c \sim \frac{e}{m} \left(\frac{\rho |\bfB|}{P_\perp}\right) R^2,
\end{equation}
and we see that it is proportional to $R^2$, when the other parameters are held constant. The Chew-Goldberger-Low EOS for $P_\perp$ \citep{CGL56} is of particular interest since the characteristic time $t_c$ and the rate $\dot{\Omega}$ are both independent of the density and the magnetic field, thereby demonstrating an unexpected universality. The resulting spin-down corresponds to the dissipation of kinetic angular momentum, which must imply that there is a corresponding increase in the intrinsic angular momentum ${\bfJ}$ (which comprises the other fluid variables).

The spin rates of low-mass stars are found to slow down by approximately two orders of magnitude over a span of $10^9$ years \citep{Sch09}. Modelling stellar spin-down is important for a multitude of reasons, including the fact that the older stars (with lower rotation rates) display lower activity in general, which has numerous ramifications for planetary habitability \citep{LL18,LL19}. We can estimate the characteristic time by choosing solar parameters (i.e., a solar-type star) for an order-of-magnitude calculation. In particular, we substitute $|\bfB|  \sim 10^{-4}$ T, $R \sim 7 \times 10^8$ m and $T \sim 5.8 \times 10^3$ K \citep{Pri14} in Eq.~(\ref{tchar}), which yields $t_c \sim 3 \times 10^6$ years. The two leading candidates invoked to explain stellar spin-down, star-disk and stellar wind braking, operate on timescales of $\sim 10^6-10^7$ years and $\sim 10^8$ years respectively \citep[Section 4.1]{BMM14}. Hence, we see that our semi-quantitative estimate is comparable to these two timescales, and may therefore constitute a viable mechanism for governing angular momentum evolution of solar-mass stars. 

The issue of angular momentum losses in protostars is another closely related topic \citep{Bod95,MP05,HHC16} which might also be resolvable through the same mechanism. We emphasize that the heuristic treatment in this subsection has primarily relied on simple scaling arguments, and a complete picture can only emerge through the synthesis of rigorous analytical models and numerical simulations. We note that this only represents the tip of the iceberg - other potential applications include pulsar braking, transport in accretion discs, and associated phenomena. In the realm of fusion, we note that the formalism developed herein may prove to be useful in explaining intrinsic rotation observed in tokamaks \citep{GDHS07,deG09,DKG13,Rice}.

\section{The Hamiltonian description and the origin of the gyromap} 
\label{Hamform}

In this section, we shall outline some of the basic principles underlying noncanonical Hamiltonian dynamics. The literature on this subject is considerable, and we refer the reader to \citet{morrison98}  for a comprehensive introduction. 

\subsection{The Lagrangian view point and the  Lagrange-Euler map}
\label{LagHamform}
First, note that the Hamiltonian can be obtained from the Lagrangian via a Legendre transform, akin to the usual process in particle mechanics. The Hamiltonian is given by
\begin{equation} \label{Hamqpi}
H[\bfq,\bfPi] = \int_D d^3a\, \dot{\bfq}\cdot \bfPi - L,
\end{equation}
where 
\bq
L[\dot{\bfq},\bfq]= \int_D d^3a\,\call\left(\bfq,\dot{\bfq},\p \bfq/\p \bfa\right)  \,,
\label{GLag}
\eq
with $L$ defined so that the action of \eqref{finalaction} is given by $S=\int_T dt\, L$.  Consequently, the  canonical momentum is given by 
\bq
\bfPi = \frac{\de L}{\de \dot \bfq}= \rho_0\dot{\bfq} + \bfPi^{\star}\,,
\label{Gmom}
\eq
and we see that we have a field theory counterpart  to the finite-dimensional case for particle motion in a magnetic field  where  the kinetic momentum differs from the canonical momentum, here with the role of the vector potential being played by $\bfPi^{\star}$. Thus \eqref{Hamqpi} gives the Hamiltonian
\bq
\label{Hamqpi2}
H=  \int_D d^3a\,\left( 
\frac{|\bfPi-\bfPi^{\star}|^2}{2\rho_0} +\rho_0 U\left(\frac{\rh_0}{\calj},\frac{\sigma_0}{\calj}\right)
+ \frac{1}{2 \calj} \q ij \, \q ik \, B_0^j B_0^k
\right)\,.
\eq
This Hamiltonian \eqref{Hamqpi2} together with the canonical Poisson bracket, 
\begin{equation} \label{canbrack}
\left\{F,G\right\} = \int_D d^3a\, \left(\frac{\p F}{\p \bfq} \cdot \frac{\p G}{\p \bfPi} - \frac{\p G}{\p \bfq} \cdot \frac{\p F}{\p \bfPi}\right), 
\end{equation}
generates the Hamiltonian equations of motion in Lagrangian variables  for our class of  three-dimensional gyroviscous magnetohydrodynamic models as follows:
\bq
\dot \bfq= \{\bfq,H\}= \frac{\de H}{\de \bfPi}\qquad \mathrm{and} \qquad \dot{\bfPi}=\{\bfPi,H\}= - \frac{\de H}{\de \bfq}\,, 
\eq
equations equivalent to the Euler-Lagrange equations obtained via $\delta S = 0$.

Now, one can use the Lagrange-Euler maps to convert both the Hamiltonian and the bracket into Eulerian variables. The procedure is described in the next section. We will see that the origin of the gyromap lies in \eqref{Gmom} and how this expression relates to different choices of Eulerian variables. The bracket obtained in terms of any of these choices  is endowed with Lie algebraic properties \citep{morrison98}, most importantly the Jacobi identity, but it does not possess the canonical form of Eq.~(\ref{canbrack}) because the Eulerian variables are not a set of canonical variables. As a result, one refers to the Hamiltonian and the bracket as being noncanonical in nature, and indeed one version is identical to that originally given in \citet{morrison80}.

As the Lagrange-Euler maps are not one-to-one, the noncanonical brackets are degenerate in general, which gives rise to the existence of invariants - the Casimirs. The theory of Casimir invariants has been studied quite extensively \citep{morrison98,morrison05,Ho08}, but there are still unresolved subtleties regarding their incompleteness, see for example \citet{yoshida14a,yoshida14b,YM16}.

The Casimirs also possess several advantages of their own, such as variational principles for Eulerian equilibria of the form
\begin{equation} \label{Feq}
\delta F := \delta (H + \lambda C) = 0,
\end{equation}
where $C$ represents any combination of all the known Casimirs. This procedure is known as the Energy-Casimir method. Once the equilibria are known, the following symmetric operator can be constructed
\begin{equation}
\Lambda_{jk}:= \frac{\delta^2 F}{\delta \psi^j \,\delta \psi^k},
\end{equation}
where $F$ is defined in Eq.~(\ref{Feq}) and the $\psi$'s denote the Eulerian (noncanonical) variables. The Energy-Casimir method states that the positive-definiteness of this operator is a sufficient condition for stability, although there are mathematical intricacies involved \citep{HMRW,rein94,batt95,morrison98,yoshida03}. Thus, the Eulerian noncanonical Hamiltonian description we obtain allows for implementation of such energy principles, although we will not pursue this application here.

\subsection{The gyro-bracket} 
\label{gyrobracket}

We shall choose our new set of observables to be the Eulerian variables  $\left\{\bfM^{c},\rho,\sigma,\bfB\right\} $,  where $\bfM^c$ was defined in  \eqref{gyroM}.  The  reason for this choice  will soon become obvious.  Recall that the Lagrange-Euler maps can be expressed in an integral form,  as they were for the density $\rho$ and and canonical momentum density $\bfM^c$ in \eqref{rhoEL} and \eqref{canMomEL}, respectively.  The remaining  Eulerian variables are given by
\begin{equation} \label{sigmaEL}
\sigma=\int_D d^{3}a\,\delta\left(\bfr-\bfq\right)\sigma_{0}(\bfa),
\end{equation}
\begin{equation} \label{magEL}
B^{j}=\int_D d^{3}a\,\delta\left(\bfr-\bfq\right)q_{,k}^{j}B_{0}^{k}(\bfa)\,. 
\end{equation}
We use these expressions  to obtain the noncanonical bracket from the canonical counterpart by the functional chain rule.  Any functional of the Eulerian observables can be  expressed  in terms of $\bfPi$ and $\bfq$; hence to  delineate, we  denote functionals of $\bfPi$ and $\bfq$ by $\bar{F}$ and those in terms of the observables by $F$, and note symbolically that $\bar{F}={F}\circ \mathfrak{E}$; consequently,
\bq
 \label{funceqv}
 \int_D d^{3}a\,\left[\frac{\delta\bar{F}}{\delta\bfPi}\cdot\delta\bfPi+\frac{\delta\bar{F}}{\delta \bfq}\cdot\delta \bfq\right] 
=\int_D d^{3}r\,\left[\frac{\delta F}{\delta \bfM^{c}}\cdot\delta \bfM^{c}+\frac{\delta F}{\delta\bfB}\cdot\delta \bfB+\frac{\delta F}{\delta\rho}\delta\rho+\frac{\delta F}{\delta\sigma}\delta\sigma\right]\,.
\eq
From Eq.~(\ref{rhoEL}), we can conclude that
\begin{equation} \label{rhorel}
\delta\rho=-\int_D d^{3}a\,\rho_{0}\nabla\delta\left(\bfr-\bfq\right)\cdot\delta \bfq,
\end{equation}
and similar identities can be found for Eqs.~\eqref{rhoEL}, \eqref{canMomEL}, \eqref{sigmaEL}, and \eqref{magEL}  as well. We substitute these identities into Eq.~(\ref{funceqv}) and carry out  integrations by parts, followed by a subsequent change in the order of integration. This results in terms that are dotted with $\delta \bfq$ and terms dotted with $\delta \bfPi$  on both the left- and right-hand sides of the expression. As $\de\bfq$ and $\de\bfPi$ are independent,  these terms must balance and thereby we obtain relationships between the Eulerian and Lagrangian functional derivatives. The algebra involved is complicated, but quite straightforward and we refer the reader to \citet{morrison09} for a more pedagogical version. The final bracket that we obtain is found to be
\bqy \label{gyroMHDbrack}
\left\{F,G\right\} &=& -\int d^{3}r\, \Bigg[M_{i}^{c}\left(\frac{\delta F}{\delta M_{j}^{c}}\partial_{j}\frac{\delta G}{\delta M_{i}^{c}}-\frac{\delta G}{\delta M_{j}^{c}}\partial_{j}\frac{\delta F}{\delta M_{i}^{c}}\right) \nonumber \\
&+& \rho\left(\frac{\delta F}{\delta M_{j}^{c}}\partial_{j}\frac{\delta G}{\delta\rho}-\frac{\delta G}{\delta M_{j}^{c}}\partial_{j}\frac{\delta F}{\delta\rho}\right)  + \sigma\left(\frac{\delta F}{\delta M_{j}^{c}}\partial_{j}\frac{\delta G}{\delta\sigma}-\frac{\delta G}{\delta M_{j}^{c}}\partial_{j}\frac{\delta F}{\delta\sigma}\right)
\nonumber \\
&+& B^{i}\left(\frac{\delta F}{\delta M_{j}^{c}}\partial_{j}\frac{\delta G}{\delta B^{i}}-\frac{\delta G}{\delta M_{j}^{c}}\partial_{j}\frac{\delta F}{\delta B^{i}}\right)
+ B^{i}\left(\frac{\delta G}{\delta B^{j}}\partial_{i}\frac{\delta F}{\delta M_{j}^{c}}-\frac{\delta F}{\delta B^{j}}\partial_{i}\frac{\delta G}{\delta M_{j}^{c}}\right)\Bigg]\,.
\eqy
By inspection, one notices that the bracket derived above is exactly the same as the three-dimensional ideal MHD bracket \citep{morrison80}; however, here  the canonical momentum $\bfM^c$  replaces the kinetic momentum $\bfM = \rho \bfv$.


Since the bracket Eq.~(\ref{gyroMHDbrack}) uses $\bfM^{c}$ as one of its observables, we must express our Hamiltonian in terms of this observable (and the others) as well. Because of the closure principle we know this is possible; indeed, \eqref{Hamqpi2} in Eulerian variables becomes
\begin{equation}
H=\int d^{3}r\,\left[\frac{\left|\bfM^{c}-\bfM^\star\right|^{2}}{2\rho}+\rho U\left(\rho,\sigma,|\bfB|\right)+\frac{|\bfB|^{2}}{2}\right]\,.
\label{HamE}
\end{equation}
The Hamiltonian of \eqref{HamE} with the bracket of \eqref{gyroMHDbrack} generates our class of three-dimensional GVMHD models in the form
\bq
\frac{\p \mathfrak{E}}{\p t}= \{\mathfrak{E},H\}\,.
\eq

On account of the fact that $\bfM = \bfM^c - \bfM^\star$, the energy has the form identical to that of  ideal MHD. This is analogous to the fact that the kinetic energy for a charged particle in a magnetic field is identical to that for a free particle. Hence, there is a choice:   one can either work with the standard ideal MHD bracket and the more complicated Hamiltonian of \eqref{HamE} in terms of the canonical momentum $\bfM^c$, or work with a complicated  bracket written in terms of the variable $\bfM$,  the conventional variable of magnetofluid theories,  and the simpler ideal MHD Hamiltonian.  To obtain the bracket in terms of $\bfM$  we  can use the gyromap \eqref{gyroM}, $\bfM = \bfM^c - \bfM^\star$,  in another chain rule calculation to transform from $\bfM^c$ to the variable $\bfM$.  This is worked out in Appendix \ref{AppB} for the case $\bfM^\star= \nabla \times \left(\calf \bfB\right)$, giving rise to a complicated Poisson bracket. 

Given that the noncanonical Poisson bracket in terms of $\bfM^c$  is the same as that of ideal MHD, it possesses the same Casimir invariants as the ideal MHD case if we replace $\bfM$  with $\bfM^c$. This use of the gyromap to obtain Casimirs, which first appeared in \citep{MCT84} and subsequently in other cases \citep{HHM87,ICTC11,ling,LM14}, differs from most of the prior studies that have sought to derive Casimirs and other conserved invariants via the HAP approach using a variety of methods, see for example \citet{morrison82,morrison98,PM96a,PM96b,hameiri,webb14a,webb14b} for a comprehensive discussion of the same. 

So, for our present general gyroviscous models, the gyromap tells us that the $\bfM$-independent Casimirs of ideal MHD will be unchanged, an example being the magnetic helicity $\int d^3r\, \bfA\cdot \bfB$. On the other hand, the cross-helicity, and other $\bfM$-dependent invariants are modified by the replacement $\bfM\rightarrow\bfM^c$.  Thus, the new cross-helicity Casimir is given by 
\bal
\int d^3r\, \frac{\bfM^c \cdot \bfB}{\rho} &= \int d^3r\, \frac{(\bfM + \bfM^{\star}) \cdot \bfB}{\rho}
\label{genhel}
\\
 &=\int d^3r\, \left[\bfv \cdot \bfB + \left(\frac{\calf}{\rho}\right) \bfB \cdot \nabla \times \bfB \right]\,.
 \label{spechel}
\eal
Equation \eqref{genhel}  is conserved for any choice of $\bfM^{\star}$ that satisfies the closure principle with a provision similar to that for conservation of the usual helicity  of MHD, viz.\ that the flow be barotropic. In \eqref{spechel} we have inserted the special case of $ \bfM^{\star}=\nabla\times  \bfL^{\star}$ with \eqref{genLform}.  The second term of \eqref{spechel} is proportional to the current helicity density, which is encountered regularly in the context of MHD \citep{Moff78,KR80,BS05,Rin19} and Hall MHD \citep{MGM02,LM15,LB16,MLB16,MaLi15,ML20} turbulence and dynamo theory.

\section{Conclusions}
 \label{Conc}
 
As we have noted in the introduction, there exist many approaches for constructing FLR models, each with their own advantages and disadvantages. In this paper, we present a HAP formalism that allows us to generate gyroviscous three-dimensional MHD models.

The action formalism allows us to clearly motivate and introduce the gyroviscous term, which is expressed in terms of a freely specifiable function. However, by using a combination of simple physical reasoning and prior results, we show that there exists a natural choice for this function, the two-dimensional limit of which exhibits consistency with the Braginskii gyroviscous tensor. We also show that the gyromap - a mathematical construct used to map back and forth between complicated Hamiltonians and easy brackets and \emph{vice versa} - emerges naturally in this framework. The HAP formalism also has the distinct advantage of generating energy-conserving models from first principles, and all our models presented conserve both energy and momentum. Through the process of reduction, we recover the noncanonical bracket for this model, and a method for finding the Casimirs is elucidated. 

One of the central results that emerged in this work was that the three-dimensional gyroviscous models do not conserve the orthodox angular momentum $\bfr \times \bfM$. We have presented a procedure for symmetrizing the momentum tensor via the construction of a hybrid momentum $\bfM^{tot}$. It is shown that the associated angular momentum $\bfr \times \bfM^{tot}$ is conserved. This procedure leads to the natural introduction of an intrinsic (spin) angular momentum which is likely to possess crucial ramifications in fusion and astrophysical plasmas; an example of the latter is briefly discussed. 

The prospects for future work are manifold. The first, and perhaps the most important from a conceptual and mathematical standpoint, is to explore the putative violation of angular momentum conservation on a Lagrangian level. The second entails the application of this framework to astrophysical and fusion systems, and thereby assess whether the ensuing results are consistent with observations. The third involves a detailed comparison with other known gyroviscous tensors, such as those formulated by \citet{braginskii65,MT71,Lil72,CS05,ramos05a,ramos05b,ramos10,ramos11,SiMo}.\footnote{The stress tensor computed by \citet{Lil72} does not rely on the large-$B$ assumption, and reduces to the Braginskii gyroviscous tensor in the small-gyroradius limit \citep{HM73}; see also \citet[Section 2.8, 2.9]{HD16}.} This is an ongoing effort, but preliminary results along this direction suggest that the symmetric part of our gyroviscous tensor might be compatible with results obtained by some of these authors, but at present we  conclude that the three-dimensional version of Branginskii's gyroviscosity tensor probably does not emerge from an action principle. A comprehensive analysis is reserved for future publications. The comparison is more tedious (albeit feasible) for the full three-dimensional case in comparison to the two-dimensional case considered in \citet{MCT84}, because the latter possessed a simple governing equation for the pressure, and it involved only two components of the momentum density and a single component of the magnetic field. 

Our model was centred on the introduction of gyroviscosity into the ideal MHD model. However, given that several variants of extended MHD possess Lagrangian and Hamiltonian formulations \citep{KLM14,LMT15,AKY15,LMM15,LMM16,LAH16,DAML,MLM,Burb}, it would seem natural to utilize the gyromap and thus formulate the gyroviscous contributions for this class of models; after doing so, their equilibria and stability can be obtained by using the HAP approach along the lines of \citet{amp0,amp1,amp2a,amp2b,ling,KTM17,KTM18,KTM20} where the stability of a variety of equilibria is analysed using Lagrangian, energy-Casimir, and dynamically accessibility methods. Likewise, this approach could also be extended to relativistic MHD and XMHD models with HAP formulations \citep{DMP15,KMM17,GTAM,CM19,Lud20}. We mention in passing that it would be interesting to explore how the  time-dependent re-gauging of \citet{amp2a} can be used to produce or remove the $\bfM^{\star}$-effects, in a manner analogous to how rotation can produce or remove effects of the magnetic field using Larmor's theorem. 

Finally we mention a most basic extension of the present work.  Our class of gyroviscous action principles  were physically motivated,  yet  ultimately {\it ad hoc}. An alternative would be to start from a more basic model, such as the Vlasov-Maxwell system, and derive a gyroviscous action by  asymptotic procedures.  A natural starting point would be the Low Lagrangian \citep{Low58} - see also \citet{MP89,morrison05} - and then reduce from phase space `fluid' element variables of that theory to the usual fluid element that we have denoted here by $\bfq(\bfa,t)$.  This would deviate from the usual historical approaches, which encompasses most of the early literature,  where one proceeds from ordering kinetic equations.  Whichever route is taken, one typically uses intuition obtained from finite-dimensional particle orbit dynamics in given strong magnetic fields, and the associated drifts, in order to make approximations, often mixing up discrete particle orbit ideas with field theoretic perturbations. It was argued in \citet{MVG13} that a more consistent approach is to remain within the field theoretic framework, and it would appear \emph{prima facie} that the Low Lagrangian is a natural framework for doing this. With this approach one could relate $\bfM^\star$ consistently to magnetization and other drifts on the fluid level. We hope to pursue this issue and others in the future. 
 
\section*{Acknowledgment}
\noindent PJM received support from the U.S. Dept.\ of Energy Contract \# DE-FG05-80ET-53088 during part of this work. AW thanks the Western New England University Research Fund for support. The authors thank the reviewers for their helpful feedback.

\appendix

\section{An Euler-Poincare approach to the three-dimensional gyromap} 
\label{AppA}

The Lagrange-Euler maps, when expressed in an integral form, are given by Eq.~(\ref{rhoEL})-(\ref{canMomEL}). Instead of ${\bfM}^c$, we can also use the velocity as our observable, and it possesses the following Lagrange-Euler map
\begin{equation} \label{velEL}
\bfv(\bfr,t)=\int_D d^{3}a\,\delta\left(\bfr-\bfq\right) \dot{\bfq}(\bfa,t) \calj,
\end{equation}
which is equivalent to $\bfv = \dot{\bfq}$, with the RHS evaluated at $\bfa=\bfq^{-1}(\bfr,t)$. The central idea is to express the Eulerian variations in terms of the Lagrangian ones, and thereby recover the equations of motion conveniently. The approach has classical roots, appeared in the plasma literature in the works of  \citet{FR60,Ka61,Low61,Lu63,Can63,Mer69,newcomb62,newcomb72,newcomb73,newcomb83}. The formalism was recast into geometric/group theoretic language in \citet{holm98}, who gave it the title of the `Euler-Poincar\'e'; this paper was motivated to a degree by what the authors called the `Arnold program' \citep{Arn66}. It should be pointed out that general variational principles of this form appeared in the early work of \citet{hamel}. The method has subsequently been applied to very many systems, including  kinetic theory \citep{cendra}, complex fluids \citep{gbr}, reduced magnetofluid models \citep{brizard10},  and hybrid fluid-kinetic models \citep{HT12,TC15,BT17,CBT18}. 

Let us illustrate this procedure by using the magnetic energy density as our example. We shall adopt the notation employed in \citet{amp2a} for convenience, where the Lagrangian displacement $\delta \bfq$ is denoted by $\bfxi$ and its Eulerianized counterpart is denoted by $\bfeta$. From Eq.~(\ref{Smag}), we know that
\begin{equation} \label{deltaSmag}
\delta S_{\mathrm{mag}} = \int_\calt dt\int_D d^3r\,\bfB\cdot\delta \bfB,
\end{equation}
where we have invoked the Eulerian closure principle. The final step lies in expressing $\delta \bfB$ in terms of $\bfeta$, which has been undertaken in \citet{FR60} (see also \citealt{amp2a}), which we list as follows:
\begin{equation}
\delta \bfB = - \nabla \times \left(\bfB \times \bfeta\right).
\end{equation}
Upon using this in Eq.~(\ref{deltaSmag}), and integrating by parts we recover the $\bfJ \times \bfB$ term, which is exactly the term arising in ideal MHD.

Upon applying the Euler-Poincar\'e method to Eq.~(\ref{finalaction}), it can be verified that one does indeed recover Eq.~(\ref{theEOM}) as our final result.

\section{The noncanonical gyroviscous bracket}
\label{AppB}

In Eq.~(\ref{gyroMHDbrack}), we presented the gyroviscous bracket in terms of the canonical momenta $M^c$ and the rest of the observables. The correspondence of the gyroviscous bracket with the ideal MHD bracket was also noted.

However, it is much more common to express noncanonical brackets in terms of the kinetic momentum $\bfM = \rho \bfv$, which we shall undertake here. In order to do so, we shall use the gyromap, discussed in Sec. \ref{Gyroact}, 
\begin{equation} \label{gyromap}
\bfM^c = \bfM + \bfM^\star = \bfM + \nabla \times \left(\calf \bfB\right),
\end{equation}
which can be easily rearranged to yield $\bfM = \bfM^c - \bfM^\star$. We shall now use the familiar concept that a given functional can be expressed in any set of (independent) observables. We denote by $F$ the functional in terms of $\bfM^c$ and the rest of the observables, and by $\tilde{F}$, the functional in terms of $\bfM$ and the rest. Since we know that $F \equiv \tilde{F}$, another chain rule calculation starts from
\bqy \label{funceqv2}
\delta F &=& \int_D d^3r\,\left[\frac{\delta F}{\delta \bfM^{c}}\cdot\delta \bfM^{c}+\frac{\delta F}{\delta \bfB}\cdot\delta \bfB+\frac{\delta F}{\delta\rho}\delta\rho+\frac{\delta F}{\delta\sigma}\delta\sigma\right] \nonumber \\
&=& \int_D d^3r\,\left[\frac{\delta \tilde{F}}{\delta \bfM}\cdot\delta \bfM+\frac{\delta \tilde{F}}{\delta \bfB}\cdot\delta \bfB+\frac{\delta \tilde{F}}{\delta\rho}\delta\rho+\frac{\delta \tilde{F}}{\delta\sigma}\delta\sigma\right] = \delta \tilde{F}, \nonumber \\
\eqy
and by using the gyromap, we find that
\begin{equation}
\delta \bfM = \delta \bfM^c - \nabla \times \left[ \delta \left(\calf \bfB\right) \right],
\end{equation}
and by substituting this into Eq.~(\ref{funceqv2}), integrating by parts and eliminating the resultant boundary terms, we finally recover the following relations
\bqy \label{gyrorels}
&&\frac{\delta F}{\delta \bfM^{c}} = \frac{\delta \tilde{F}}{\delta \bfM}, \nonumber \\
&&\frac{\delta F}{\delta\rho} = \frac{\delta \tilde{F}}{\delta\rho} - \bfB\cdot \left(\nabla \times \frac{\delta \tilde{F}}{\delta \bfM}\right) \frac{\p \calf}{\p \rh}, \nonumber \\
&&\frac{\delta F}{\delta\sigma} = \frac{\delta \tilde{F}}{\delta\sigma} - \bfB\cdot \left(\nabla \times \frac{\delta \tilde{F}}{\delta \bfM}\right) \frac{\p \calf}{\p \sigma}, \nonumber \\
&&\frac{\delta F}{\delta \bfB} = \frac{\delta \tilde{F}}{\delta \bfB} - \left[\bfB\cdot \left(\nabla \times \frac{\delta \tilde{F}}{\delta \bfM}\right)\right] \frac{\p \calf}{\p |\bfB|} \frac{\bfB}{|\bfB|}  - \left(\nabla \times \frac{\delta \tilde{F}}{\delta \bfM}\right) \calf.
\eqy
We can now recover the bracket in terms of $\bfM$ from Eq.~(\ref{gyroMHDbrack}), by implementing the following two successive steps.
\begin{enumerate}
\item First, replace the $M^c_i$ in the first line of Eq.~(\ref{gyroMHDbrack}), prior to the functional derivatives, with Eq.~(\ref{gyromap}). This ensures that only $\bfM$ and the other observables are present.
\item Next, the functional derivatives occurring in Eq.~(\ref{gyroMHDbrack}) should be replaced with the relations delineated in Eq.~(\ref{gyrorels}).
\end{enumerate}
We shall not list the final bracket in its entirety since its complexity is clearly self-evident\footnote{An explicit example of the gyro-bracket in terms of $\bfM$ for a particular choice of the gyromap was presented in A.~Wurm and P.~J.~Morrison,
Derivation of Hamiltonian magnetofluid models with gyroviscous-like contributions using a gyromap, Sherwood International Fusion Theory conference, Santa Fe, NM, April 15-17 (2013).}. Hence, this illustrates the advantage of the gyromap in facilitating a much simpler bracket. Simply through the process of inspection, it would have been almost impossible to construct the bracket in terms of $\bfM$ or to find the variable $\bfM^c$ that simplified the bracket.

The Hamiltonian, in terms of $\bfM$, is much simpler as seen from the following expression.
\begin{equation}
H=\int d^{3}r\,\left[\frac{|\bfM|^{2}}{2\rho}+\rho U\left(\rho,\sigma\right)+\frac{|\bfB|^{2}}{2}\right].
\end{equation}
In other words, the resultant Hamiltonian is exactly identical to the total energy associated with ideal MHD \citep{morrison80,Fre14,GKP19}.

We note, that any choice for $\bfM^{\star}$ that satisfies the Eulerian closure principle will, under an analogous transformation, yield a complicated bracket in terms of $\bfM$, yet one that reduces  to the MHD bracket of \citet{morrison80} when the variable $\bfM^c$ is used. Thus, for any choice we have the same trade-off between Hamiltonian and bracket.


\end{document}